# Collider Implications of a Non-Universal Higgs

**C.D. McMullen[a*] and S. Nandi[b†]**

[a] *Department of Physics, Louisiana School for Math, Science, and the Arts*
*Natchitoches, LA 71457, USA*

[b] *Department of Physics, Oklahoma State University*
*Stillwater, OK 74078, USA*

**Abstract**

We consider in detail the possibility that the Higgs is wholly or partially excluded from propagating into one otherwise universal extra dimension. This exclusion of the Higgs from propagating into an otherwise universal extra dimension violates tree-level Kaluza-Klein number conservation in the Yukawa interactions. As a consequence, there is inter-mode mixing between fermions. For example, zero-mode fermions mix with their associated Kaluza-Klein excitations. This is in contrast to the original universal extra dimensions scenario, in which conservation of Kaluza-Klein number prohibits such inter-mode mixing. Inter-mode mixing is especially significant for the top quark, since its mass (171.4 ± 2.1 GeV) is approximately one-half the current Tevatron mass bound (~350-400 GeV) for Kaluza-Klein excitations of quarks propagating into universal extra dimensions. We compute the effects that mixing among the zero-modes and lowest-lying Kaluza-Klein excitations has on the lightest third-generation charge 2/3 quark mass eigenvalue in the non-universal Higgs model with one otherwise universal extra dimension. Another consequence of the inter-mode mixing is that the Kaluza-Klein excitations of the fermions can decay to a zero-mode Higgs and a corresponding zero-mode fermion. As a result, the pair production of Kaluza-Klein excitations of the top quark would lead to two zero-mode Higgs bosons plus a zero-mode top quark/anti-quark pair. We compute the cross section that the non-universal Higgs model contributes to Higgs production at the Large Hadron Collider. The effect is quite large: For example, the Kaluza-Klein contribution to Higgs production is comparable to or larger than the Standard Model contribution, depending on the Higgs mass, for compactification scales up to about 600 GeV.

* email: cmcmullen@lsmsa.edu
† email: s.nandi@okstate.edu

## 1. Introduction

We investigate the possibility that the Standard Model (SM) fermions and gauge bosons propagate into one large ($\mathrm{TeV}^{-1}$-size) [1] extra dimension, while the Higgs is wholly or partially excluded from propagating into the otherwise universal extra dimension (OUED). Such a non-universal Higgs scenario has some phenomenologically interesting features compared to the original universal extra dimensions (UED) model [2]. One motivation for a non-universal Higgs is that there is a little hierarchy problem in the case of a UED bulk Higgs [3], which pushes the mass of the zero mode Higgs boson to the 5D cut off scale; the problem gets worse in six or higher dimensions. An immediate consequence of a non-universal Higgs is that Kaluza-Klein (KK) number is not conserved in the tree-level Yukawa interactions. This KK number violation results in inter-mode quark mixing – i.e. there is mixing between different levels of the associated KK excitations of the quarks. This is in contrast to the original UED model, in which there is only mixing between KK excitations of the same level. Inter-mode quark mixing is especially significant for the top quark because the top quark mass (171.4 ± 2.1 GeV) [4] is approximately one-half the current Tevatron mass bound (~350-400 GeV) for KK excitations of quarks propagating into one UED [5],[6]. We find that the lightest third-generation charge $2/3$ quark mass eigenvalue is very sensitive to inter-mode quark mixing in the non-universal Higgs model.

Another phenomenological difference between the non-universal Higgs and UED models lies in the decay mechanism for the KK excitations. In the UED model, conservation of KK number prohibits tree-level decays of the KK excitations. However, the heavier KK modes may decay to the lightest KK particle (LKP) via radiative corrections [6] to the masses of the KK excitations. The fat brane scenario provides an alternative decay mechanism: The SM fields reside on a fat brane of radius $R \sim \mathrm{TeV}^{-1}$, which is much smaller than the size $r$ of the extra dimension [7]. While the SM fields have associated KK excitations with masses on the order of a TeV, the KK excitations of the gravitons can be as light as ~sub-$\mathrm{mm}^{-1}$. Thus, the KK excitations of the SM fields can decay to a corresponding zero-mode and a KK excitation of a graviton. One striking feature of the non-universal Higgs model is that the KK excitations of the quarks can decay to the corresponding zero-mode quark and a Higgs at tree-level. This decay mode is especially dominant for the KK excitation of the top quark where this mixing is large. This is a very important consequence of KK number violation in the Yukawa interactions.

Our paper is organized as follows. In Sec. 2, we develop the inter-mode quark mixing matrix from the Yukawa interactions for the quarks in the effective 4D theory for the non-universal Higgs model. We discuss this inter-mode quark mixing in Sec. 3, where we compute the effects that mixing among the zero-mode and lowest-lying Kaluza-Klein excitations has in the top sector. In Sec. 4, we consider the collider phenomenology of the non-universal Higgs model. In particular, we compute the cross section for the pair production of KK excitations of the top quark at the Large Hadron Collider (LHC), and discuss the subsequent decay of the KK excitation of the top quark to $tH$ giving rise to $t\bar{t}HH$. This will be a very important new source of Higgs boson production at the LHC. We draw our conclusions in Sec. 5.

## 2. Yukawa Interactions

The Higgs is non-universal if it is wholly or partially excluded from propagating into one $\text{TeV}^{-1}$-size OUED. If the Higgs is wholly excluded, this does not necessarily mean that the Higgs does not propagate into any extra dimensions: It could simply be that the Higgs propagates into a different $\text{TeV}^{-1}$-size extra dimension. In this case, the effective 4D Lagrangian density for the Yukawa interactions of the 5D quark fields is

$$L^Y(x^\mu) = \int_{y=0}^{\pi R} \int_{y_H=0}^{\pi R_H} \left\{ \sum_{i,j=1}^{3} \overline{Q}_i(x^\mu, y)\left[\widetilde{y}_5^{ij} U_j(x^\mu, y)\delta(y_H)\widetilde{\Phi}(x^\mu, y_H)\delta(y) + y_5^{ij} D_j(x^\mu, y)\delta(y_H)\Phi(x^\mu, y_H)\delta(y)\right] + \text{h.c}\right\} dy_H dy \tag{1}$$

where $\{x^\mu\}$ are the usual 4D spacetime coordinates, $y$ and $y_H$ are the extra-dimensional coordinates corresponding to the quarks and the Higgs, respectively, the indices $\{i, j\} \in \{1,2,3\}$ represent the quark generations, $\{\widetilde{y}_5^{ij}, y_5^{ij}\}$ are the 5D Yukawa couplings, $\Phi(x^\mu, y_H)$ represents the 5D Higgs doublet, $\widetilde{\Phi}(x^\mu, y_H) = i\tau_2\Phi^*(x^\mu, y_H)$ is its conjugate multiplet, and the 5D quark multiplets $Q_i(x^\mu, y)$, $U_i(x^\mu, y)$, and $D_i(x^\mu, y)$ consist of four-component vector-like quark fields. The delta functions $\delta(y)$ and $\delta(y_H)$ prohibit the Higgs and quarks, respectively, from propagating into the other extra dimension.

As an alternative to Eq. 1, the Higgs may be partially excluded $(0 < R_H < R)$ from propagating into one OUED. That is, the Higgs is constrained to lie within $0 \le y \le R_H$, whereas the extra-dimensional coordinate $y$ is unrestrained for the other SM fields. This bears similarity to a fat brane scenario [7]. The original UED fat brane model distinguished between the gravitons, which had full freedom in an extra dimension of radius $r$, and the SM fields, which resided on a fat brane of thickness $R << r$ in the same extra dimension. However, in the non-universal Higgs model the SM fields have full freedom in an extra dimension of radius $R$ while the Higgs resides on a subset $0 \le y \le R_H$ of the extra dimension. In this case, the effective 4D Lagrangian density for the Yukawa interactions of the 5D quark fields is

$$L^Y(x^\mu) = \int_{y=0}^{\pi R_H} \left\{ \sum_{i,j=1}^{3} \overline{Q}_i(x^\mu, y)\left[\widetilde{y}_5^{ij} U_j(x^\mu, y)\widetilde{\Phi}(x^\mu, y) + y_5^{ij} D_j(x^\mu, y)\Phi(x^\mu, y)\right] + \text{h.c.}\right\} dy \tag{2}$$

where the upper limit of the $y$-integration accounts for the fact that the Higgs does not propagate into the entirety of the compactified dimension $(0 < R_H < R)$.

Henceforth in our analysis, we focus on the case of partial exclusion of the Higgs (Eq. 2), since the case where the Higgs is wholly excluded corresponds to the limit that $R_H \to 0$ (Eq. 1). The 5D quark multiplets can be decomposed into 4D two-component Weyl spinors via Fourier expansion about the compactified coordinate $y$:

$$
\begin{aligned}
Q_i\left(x^\mu, y\right) &= \frac{1}{\sqrt{\pi R}}\left\{q_{iL}^{(0)}\left(x^\mu\right) + \sqrt{2}\sum_{n=1}^{\infty}\left[q_{iL}^{(n)}\left(x^\mu\right)\cos\left(\frac{ny}{R}\right) + q_{iR}^{(n)}\left(x^\mu\right)\sin\left(\frac{ny}{R}\right)\right]\right\} \\
U_i\left(x^\mu, y\right) &= \frac{1}{\sqrt{\pi R}}\left\{u_{iR}^{(0)}\left(x^\mu\right) + \sqrt{2}\sum_{n=1}^{\infty}\left[u_{iR}^{(n)}\left(x^\mu\right)\cos\left(\frac{ny}{R}\right) + u_{iL}^{(n)}\left(x^\mu\right)\sin\left(\frac{ny}{R}\right)\right]\right\} \qquad (3) \\
D_i\left(x^\mu, y\right) &= \frac{1}{\sqrt{\pi R}}\left\{d_{iR}^{(0)}\left(x^\mu\right) + \sqrt{2}\sum_{n=1}^{\infty}\left[d_{iR}^{(n)}\left(x^\mu\right)\cos\left(\frac{ny}{R}\right) + d_{iL}^{(n)}\left(x^\mu\right)\sin\left(\frac{ny}{R}\right)\right]\right\}
\end{aligned}
$$

where half of the zero modes, which are not observed in 4D, have been projected out via a simple orbifold compactification choice – namely, $S^1/Z_2$ $\left(Z_2 : y \to -y\right)$. Associated with the SM quark doublet $q_{iL}^{(0)}\left(x^\mu\right) \equiv \begin{pmatrix} u_i^{(0)} \\ d_i^{(0)} \end{pmatrix}_L$ are left-chiral and right-chiral KK doublets $q_{iL}^{(n)}\left(x^\mu\right)$ and $q_{iR}^{(n)}\left(x^\mu\right)$, and corresponding to the SM quark singlets $u_{iR}^{(0)}\left(x^\mu\right)$ and $d_{iR}^{(0)}\left(x^\mu\right)$ are KK singlets $u_{iL}^{(n)}\left(x^\mu\right)$, $u_{iR}^{(n)}\left(x^\mu\right)$, $d_{iL}^{(n)}\left(x^\mu\right)$, and $d_{iR}^{(n)}\left(x^\mu\right)$. The 5D Higgs doublet is even under the orbifold transformation $\Phi\left(x^\mu, -y\right) = \Phi\left(x^\mu, y\right)$ in order to obtain the usual SM Higgs doublet in the effective 4D theory:

$$
\Phi\left(x^\mu, y\right) = \frac{1}{\sqrt{\pi R_H}}\left[\Phi^{(0)}\left(x^\mu\right) + \sqrt{2}\sum_{n=1}^{\infty}\Phi^{(n)}\left(x^\mu\right)\cos\left(\frac{ny}{R_H}\right)\right] \qquad (4)
$$

We assume that only the zero mode of the Higgs acquires a vev, corresponding to the usual SM Higgs vev $\upsilon = \left(\sqrt{2}G_F\right)^{-1} = 247\,\text{GeV}$ [8].

Mixing between SM quarks and their associated KK excitations is most significant for the third generation since the top quark mass (171.4 ± 2.1 GeV) is approximately one-half the current Tevatron mass bound (~350-400 GeV) for KK excitations of quarks propagating into one UED. First and second generation quark mixing will be comparatively less significant. We define the effective third-generation 4D Yukawa couplings in terms of their corresponding 5D couplings as

$$
\tilde{y}_4^{33} = \frac{\tilde{y}_5^{33}}{R}\sqrt{\frac{R_H}{\pi}} \quad , \quad y_4^{33} = \frac{y_5^{33}}{R}\sqrt{\frac{R_H}{\pi}} \qquad (5)
$$

In this model, the effective 4D Yukawa interactions among the SM zero mode quarks are

$$
L_{33}^{(0,0)}\left(x^\mu\right) = \bar{q}_{3L}^{(0)}\left(x^\mu\right)\left[\tilde{y}_4^{33}u_{3R}^{(0)}\left(x^\mu\right) + y_4^{33}d_{3R}^{(0)}\left(x^\mu\right)\right]\frac{\upsilon + h}{\sqrt{2}} + \text{h.c.} \qquad (6)
$$

while those among the KK excitations of the quarks of the same mode are

$$L_{33}^{(n,n)}\left(x^{\mu}\right)=\sum_{n=1}^{\infty}\left\{\left[1+\frac{R}{2\pi nR_H}\sin\left(\frac{2\pi nR_H}{R}\right)\right]\bar{q}_{3L}^{(n)}\left(x^{\mu}\right)\left[\tilde{y}_4^{33}u_{3R}^{(n)}\left(x^{\mu}\right)+y_4^{33}d_{3R}^{(n)}\left(x^{\mu}\right)\right]\right.$$
$$\left.+\left[1-\frac{R}{2\pi nR_H}\sin\left(\frac{2\pi nR_H}{R}\right)\right]\bar{q}_{3R}^{(n)}\left(x^{\mu}\right)\left[\tilde{y}_4^{33}u_{3L}^{(n)}\left(x^{\mu}\right)+y_4^{33}d_{3L}^{(n)}\left(x^{\mu}\right)\right]\right\}\frac{\upsilon+h}{\sqrt{2}}+\text{h.c.} \tag{7}$$

The 4D effective Yukawa interaction between the SM quarks and their KK excitations are given by

$$L_{33}^{(0,n)}\left(x^{\mu}\right)=\sum_{n=1}^{\infty}\frac{R}{\pi nR_H}\sin\left(\frac{\pi nR_H}{R}\right)\bar{q}_{3L}^{(0)}\left(x^{\mu}\right)\left[\tilde{y}_4^{33}u_{3R}^{(n)}\left(x^{\mu}\right)+y_4^{33}d_{3R}^{(n)}\left(x^{\mu}\right)\right]\frac{\upsilon+h}{\sqrt{2}}+\text{h.c.} \tag{8}$$

with similar expressions for $L_{33}^{(n,0)}$. In addition, there are Yukawa interactions between KK excitations of quarks of different modes.

The fermions receive mass contributions from the vev's of the SM Higgs doublet $\Phi^{(0)}\left(x^{\mu}\right)$ and conjugate doublet $\tilde{\Phi}^{(0)}\left(x^{\mu}\right)$ as well as $\pm n/R$ contributions from the kinetic terms for the KK excitations. The truncated mass matrix for the SM top quark and its first KK excitations is

$$\begin{pmatrix}\bar{q}_{3L}^{(0)} & \bar{q}_{3L}^{(1)} & \bar{u}_{3L}^{(1)}\end{pmatrix}\begin{pmatrix}\frac{\tilde{y}_4^{33}\upsilon}{\sqrt{2}} & 0 & \frac{\tilde{y}_4^{33}\upsilon R}{\pi R_H}\sin\left(\frac{\pi R_H}{R}\right)\\ \frac{\tilde{y}_4^{33}\upsilon R}{\pi R_H}\sin\left(\frac{\pi R_H}{R}\right) & \pm\frac{1}{R} & \frac{\tilde{y}_4^{33}\upsilon}{\sqrt{2}}\left[1+\frac{R}{2\pi R_H}\sin\left(\frac{2\pi R_H}{R}\right)\right]\\ 0 & \frac{\tilde{y}_4^{33}\upsilon}{\sqrt{2}}\left[1-\frac{R}{2\pi R_H}\sin\left(\frac{2\pi R_H}{R}\right)\right] & \mp\frac{1}{R}\end{pmatrix}\begin{pmatrix}u_{3R}^{(0)}\\ q_{3R}^{(1)}\\ u_{3R}^{(1)}\end{pmatrix} \tag{9}$$

The $(1,2)$ and $(3,1)$ elements are zero in accordance with $SU(2)$ invariance. The corresponding matrix for the Yukawa interactions will not have the $\pm 1/R$ terms in the in the diagonal $(2,2)$ and $(3,3)$ elements because these contributions come from the 5D kinetic terms. Thus, we see that the mass matrix and the Yukawa coupling matrix are not proportional. So when we diagonalize the mass matrix, the Yukawa coupling matrix will not be diagonal, leading to off diagonal Yukawa interactions between the top quark, its KK excitations, and the Higgs boson. There is a similar truncated mass matrix for the SM bottom quark and its first KK excitations:

$$\begin{pmatrix}\bar{q}_{3L}^{(0)} & \bar{q}_{3L}^{(1)} & \bar{d}_{3L}^{(1)}\end{pmatrix}\begin{pmatrix}\frac{y_4^{33}\upsilon}{\sqrt{2}} & 0 & \frac{y_4^{33}\upsilon R}{\pi R_H}\sin\left(\frac{\pi R_H}{R}\right)\\ \frac{y_4^{33}\upsilon R}{\pi R_H}\sin\left(\frac{\pi R_H}{R}\right) & \pm\frac{1}{R} & \frac{y_4^{33}\upsilon}{\sqrt{2}}\left[1+\frac{R}{2\pi R_H}\sin\left(\frac{2\pi R_H}{R}\right)\right]\\ 0 & \frac{y_4^{33}\upsilon}{\sqrt{2}}\left[1-\frac{R}{2\pi R_H}\sin\left(\frac{2\pi R_H}{R}\right)\right] & \mp\frac{1}{R}\end{pmatrix}\begin{pmatrix}d_{3R}^{(0)}\\ q_{3R}^{(1)}\\ d_{3R}^{(1)}\end{pmatrix} \tag{10}$$

However, this inter-mode mixing in the bottom sector, as well as for the light quark sector, will be very small because of the tiny masses of these quarks compared to the compactification scale.

In the limit that $R_H \to R$, the SM quark mixing decouples from the KK quark mixing. There is no inter-mode mixing in this extreme. The limit that $R_H \to 0$ corresponds to the case where the Higgs is wholly excluded (Eq. 1). Inter-mode mixing is maximal in this case. In this case, the truncated mixing matrix for the SM top quark and its first KK excitations simplifies to

$$\begin{pmatrix} \bar{q}_{3L}^{(0)} & \bar{q}_{3L}^{(1)} & \bar{u}_{3L}^{(1)} \end{pmatrix} \begin{pmatrix} \frac{\tilde{y}_4^{33}\upsilon}{\sqrt{2}} & 0 & \tilde{y}_4^{33}\upsilon \\ \tilde{y}_4^{33}\upsilon & \pm\frac{1}{R} & \tilde{y}_4^{33}\upsilon\sqrt{2} \\ 0 & 0 & \mp\frac{1}{R} \end{pmatrix} \begin{pmatrix} u_{3R}^{(0)} \\ q_{3R}^{(1)} \\ u_{3R}^{(1)} \end{pmatrix} \qquad (11)$$

## 3. Inter-Mode Quark Mixing

Since the top quark mass may be as large as roughly one-half the compactification scale, inter-mode quark mixing is most significant for the third-generation charge 2/3 quarks. The $n \geq 2$ KK excitations will be at least four times as heavy as the zero-mode. Thus, the $n \geq 2$ KK modes approximately decouple compared to the mixing between the zero-mode and the $n = 1$ KK modes. Therefore, we restrict our attention to the $3 \times 3$ mixing among $n = 0$ and $n = 1$ charge $2/3$ third generation quarks.

Recall that for a given mode $n$ there are twice as many KK excitations as zero modes due to the Fourier expansions of the 5D quark fields (Eq. 3). For example, the first generation of zero modes includes the left-chiral doublet $q_{1L}^{(0)}$ and the right-chiral singlets $u_{1R}^{(0)}$ and $d_{1R}^{(0)}$, while the associated $n = 1$ KK excitations include the left-chiral doublet $q_{1L}^{(1)}$, the right-chiral doublet $q_{1R}^{(1)}$, the right-chiral singlets $u_{1R}^{(1)}$ and $d_{1R}^{(1)}$, and the left-chiral singlets $u_{1L}^{(1)}$ and $d_{1L}^{(1)}$. We emphasize that $u_{1L}^{(1)}$ and $d_{1L}^{(1)}$ are left-chiral singlets from the Fourier expansions of the 5D quark singlets $U(x^\mu, y)$ and $D(x^\mu, y)$ – not to be confused with the charge $2/3$ and charge $-1/3$ components of the quark doublet $q_{1L}^{(1)}$. In order to distinguish between the left-chiral singlets $u_{1L}^{(1)}$ and $d_{1L}^{(1)}$ and the charge $2/3$ and charge $-1/3$ components of the quark doublet $q_{1L}^{(1)}$, we use the same symbol $q_{1L}^{(1)}$ to represent the doublet as well as its charge $2/3$ and charge $-1/3$ components, where the distinction should be clear from the context.

The observed left handed top quark, with a mass of 171.4 ± 2.1 GeV, is actually not the zero mode of the top quark $u_3^{(0)}$ in Eq. 3, but a linear combination of $\{q_3^{(0)}, q_3^{(1)}, u_3^{(1)}\}$. The mass eigenstates $\{t, t_1^{\bullet}, t_1^{\circ}\}$ are related to the weak eigenstates $\{q_3^{(0)}, q_3^{(1)}, u_3^{(1)}\}$ via a bi-unitary transformation

$$
\begin{aligned}
\left(t, t_1^{\bullet}, t_1^{\circ}\right)_L^T &= U_L^{-1}\left(q_3^{(0)}, q_3^{(1)}, u_3^{(1)}\right)_L^T \\
\left(t, t_1^{\bullet}, t_1^{\circ}\right)_R^T &= U_R^{-1}\left(u_3^{(0)}, q_3^{(1)}, u_3^{(1)}\right)_R^T
\end{aligned}
\tag{12}
$$

which can be expressed in the form

$$
\begin{aligned}
t_L &= c_a q_{3L}^{(0)} + s_a c_b q_{3L}^{(1)} + s_a s_b u_{3L}^{(1)} \\
t_{1L}^{\bullet} &= -s_a c_c q_{3L}^{(0)} + \left(c_a c_b c_c - s_b s_c\right) q_{3L}^{(1)} + \left(c_a s_b c_c + c_b s_c\right) u_{3L}^{(1)} \\
t_{1L}^{\circ} &= -s_a s_c q_{3L}^{(0)} + \left(c_a c_b s_c + s_b c_c\right) q_{3L}^{(1)} + \left(c_a s_b s_c - c_b c_c\right) u_{3L}^{(1)}
\end{aligned}
\tag{13}
$$

with analogous expressions for the right-chiral quark fields. The $\{s_i, c_i\}$ are short for $\{\sin\theta_i, \cos\theta_i\}$.

In the case where the Higgs is wholly excluded from the OUED (Eq. 11), there are two unknown parameters involved in the $3\times 3$ top quark mixing: $\tilde{y}_4^{33}$ and $1/R$. The observed top quark mass $m_t = 171.4 \pm 2.1$ GeV, being the lightest eigenvalue of the top quark mass matrix $M_t$, constrains $\tilde{y}_4^{33}$ for a given value of $1/R$, and vice-versa. Just one additional unknown parameter, $y_4^{33}$, is involved in the $3\times 3$ bottom quark mass matrix $M_b$, which is compensated for by fact that the lightest eigenvalue of the bottom quark mass matrix $M_b$ equals the observed bottom quark mass $m_b$. There is one additional parameter, $R_H / R$, in the case where the Higgs is partially excluded from the OUED (Eq. 9).

We illustrate the $3\times 3$ mixing among $n=0$ and $n=1$ charge $2/3$ third generation quarks with the following numerical example: $R = (500\,\text{GeV})^{-1}$, $\tilde{y}_4^{33} = 1.35$, and $R_H = 0$ (corresponding to the case where the Higgs is wholly excluded from the OUED). In this case, the $3\times 3$ top quark mass matrix assumes the form

$$
M_t\left(1/R = 500\,\text{GeV}, \tilde{y}_4^{33} = 1.35\right) = \begin{pmatrix} 236 & 0 & 333 \\ 333 & 500 & 472 \\ 0 & 0 & -500 \end{pmatrix}
\tag{14}
$$

We diagonalize $M_t M_t^{\dagger}$ to find the corresponding mass-squared eigenvalues:

$$
m_t = 172.3\,\text{GeV} \quad ; \quad m_{t_1^{\bullet}}, m_{t_1^{\circ}} = 376\,\text{GeV}, 911\,\text{GeV}
\tag{15}
$$

The unitary matrices, $U_L$ and $U_R$, are found by diagonalizing $M_t M_t^{\dagger}$ and $M_t^{\dagger} M_t$, respectively:

$$U_L = \begin{pmatrix} 0.397 & 0.863 & 0.311 \\ 0.804 & -0.164 & -0.571 \\ -0.442 & 0.477 & -0.760 \end{pmatrix} \quad , \quad U_R = \begin{pmatrix} 0.397 & 0.863 & 0.311 \\ 0.442 & -0.477 & 0.760 \\ 0.804 & -0.164 & -0.571 \end{pmatrix} \tag{16}$$

We wrote a Fortran program to carry out the bi-unitary transformation numerically. The algorithm diagonalizes a $3\times3$ matrix via the brute force solution to the characteristic cubic. The routine includes numerous checks to verify – using a small tolerance in double precision – that the eigenvalues satisfy the eigenvalue equation, that the unitary matrices have orthonormal rows and columns, and that the eigenvalues and corresponding unitary matrices satisfy the linear algebra associated with the bi-unitary transformation.

It is useful to define the third-generation charge $2/3$ quark Yukawa coupling in terms of a new coupling as follows:

$$\tilde{\alpha}_4^{33} \equiv \frac{\left(\tilde{y}_4^{33}\right)^2}{4\pi} \tag{17}$$

This coupling is perturbative if it is less than unity $\left(\tilde{\alpha}_4^{33} < 1\right)$. We will refer to this modified third-generation charge $2/3$ quark Yukawa coupling $\tilde{\alpha}_4^{33}$ in the ensuing analysis.

In the case where the Higgs is wholly excluded from the OUED (i.e. $R_H = 0$), the two independent model parameters include the modified third-generation charge $2/3$ quark Yukawa coupling $\tilde{\alpha}_4^{33}$ and the quark compactification scale $1/R$. Fig. 1 shows the allowed region of parameter space for the case $R_H = 0$ at the 1-sigma (black) and 3-sigma (gray) level. Fig. 2 illustrates the mass eigenvalues $m_t, m_{t_1^\bullet}$, and $m_{t_1^\circ}$ as functions of the quark compactification scale $1/R$ for the central allowed values of $\tilde{\alpha}_4^{33}$ from Fig. 1. The inter-mode quark mixing is also responsible for the mass splitting between $m_{t_1^\bullet}$ and $m_{t_1^\circ}$. The effect of the inter-mode quark mixing is quite large. A smaller $\tilde{\alpha}_4^{33}$ demands a greater compactification scale $1/R$, and a smaller $1/R$ necessitates a greater $\tilde{\alpha}_4^{33}$. A smaller quark compactification scale $1/R$ or a greater coupling $\tilde{\alpha}_4^{33}$ results in greater splitting between the KK masses $m_{t_1^\bullet}$ and $m_{t_1^\circ}$ (upper black and white curves with similarly shaped markers).

In the case where the Higgs is partially excluded from the OUED $\left(0 < R_H < R\right)$, the three independent model parameters include the modified third-generation charge $2/3$ quark Yukawa coupling $\tilde{\alpha}_4^{33}$, the quark compactification scale $1/R$, and the ratio of the Higgs radius to the quark radius $R_H / R \le 1$. If $\tilde{\alpha}_4^{33}$ is on the order of its SM value, $m_t^2 / 2\pi\upsilon^2 \approx 1/4\pi$, then either the quark compactification scale must be very large or the

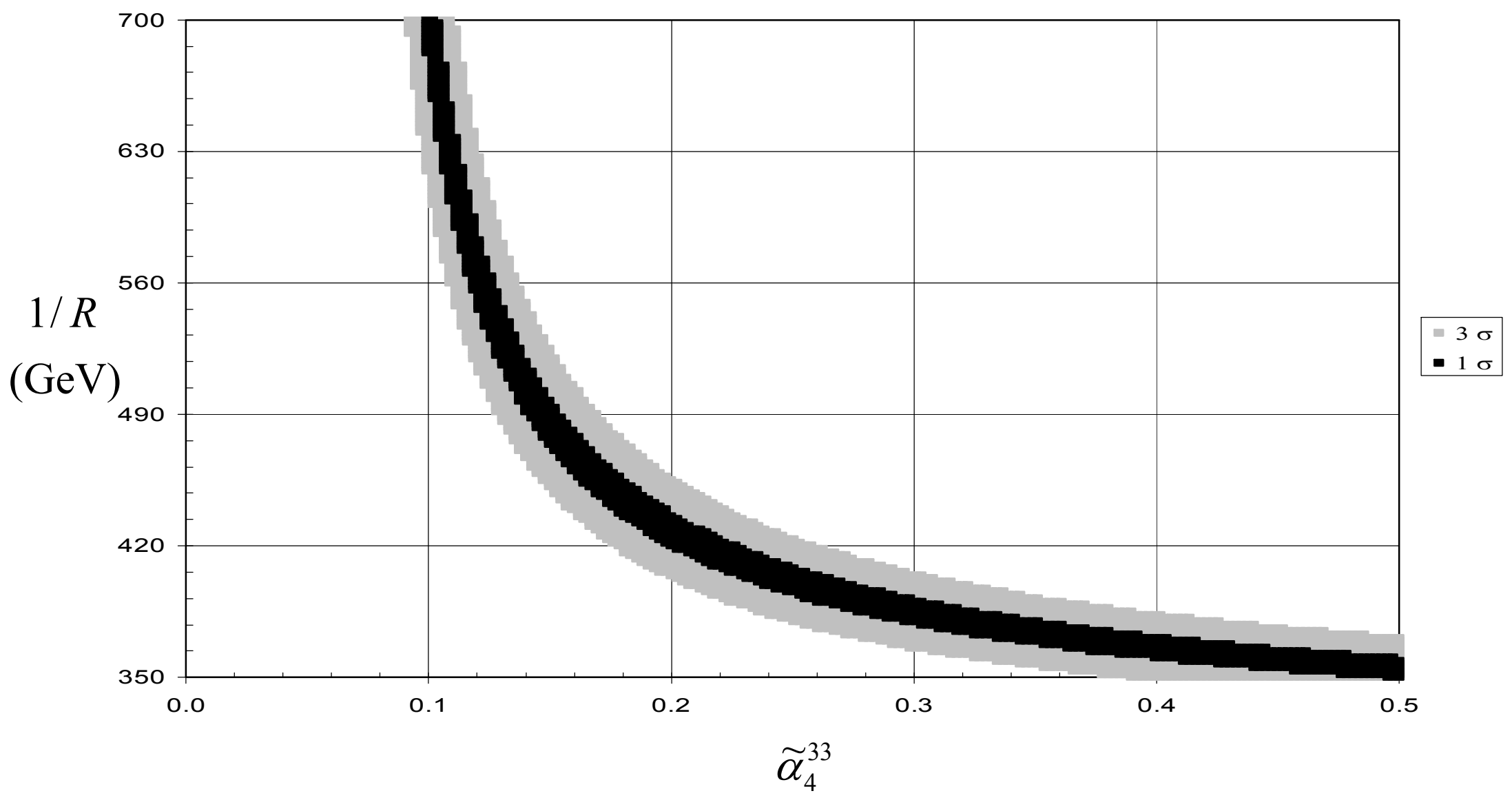

**Fig. 1**. The allowed ranges of the coupling $\tilde{\alpha}_4^{33}$ and quark compactification scale $1/R$ are depicted for the case $R_H = 0$ at the 1-sigma (■) and 3-sigma (■) level, using $m_t = 171.4 \pm 2.1\,\mathrm{GeV}$.

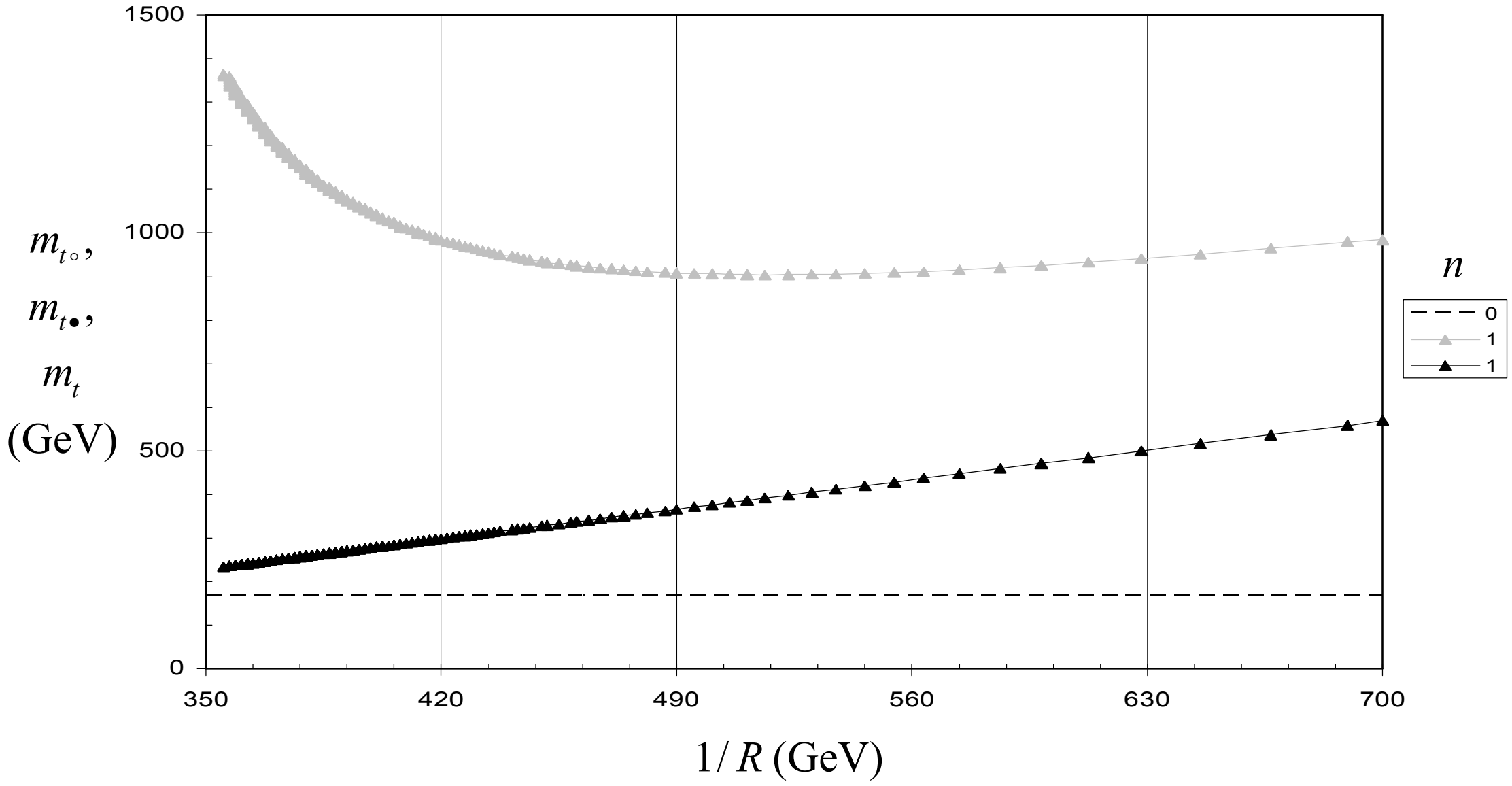

**Fig. 2.** The mass eigenvalues $m_t, m_{t_1^\bullet}$, and $m_{t_1^\circ}$ of the top quark mass matrix $M_t$ are illustrated as functions of the quark compactification scale $1/R$ for the central allowed values of $\tilde{\alpha}_4^{33}$ from Fig. 1. The black (▲) and gray (▲) curves illustrate the mass splitting between $m_{t_1^\bullet}$ and $m_{t_1^\circ}$. The dashed line (– – –) depicts the observed top quark mass $m_t = 171.4 \pm 2.1\,\mathrm{GeV}$.

Higgs must have nearly full freedom in the quarks' extra dimension (i.e. $R_H / R \approx 1$). As the ratio $R_H / R$ approaches unity, the KK quarks decouple from the SM and the KK masses become degenerate. In the other extreme, as $R_H$ approaches zero, the inter-mode quark mixing becomes maximal.

Fig. 3 shows the allowed region of parameter space for the case $0 < R_H < R$ at the 1-sigma (black) and 3-sigma (gray) level for three different values of $\tilde{\alpha}_4^{33}$. The case $\tilde{\alpha}_4^{33} = 0.077$ corresponds to its SM value $\approx 1/4\pi$. In this case, the allowed region adjoins the right edge of the graph where $R_H \to R$, where the SM quarks decouple from their associated KK excitations. As $\tilde{\alpha}_4^{33}$ increases, the allowed region becomes narrower and moves toward the left, where $R_H$ is smaller and inter-mode quark mixing is enhanced. Fig. 4 illustrates the mass eigenvalues $m_t, m_{t_1^\bullet}$, and $m_{t_1^\circ}$ as functions of the quark compactification scale $1/R$ for the central allowed values of $R_H / R$ from Fig. 3.

We now discuss how to generalize the inter-mode quark mixing to include the first and second generations of quarks and higher-level KK excitations. The complete inter-mode quark mixing is described by $(6N+3)\times(6N+3)$ mass matrices, which involve 20 independent unknown parameters: $\{\tilde{y}_4^{ij}, y_4^{ij}\}$, $1/R$, and $R_H / R$. These 20 unknown parameters are immediately constrained by the 6 observed quark masses and the 9 elements of the KM matrix. Here, $N \equiv n_{\max}$ represents the number of KK levels present in the theory – i.e. the sums in Eqs. 3-4 are not actually infinite in practice, but truncated such that $n_{\max} / R$ does not exceed the cut-off scale of the 4D theory.

In generation space, the mass matrices $M_u$ and $M_u^\dagger$ for the charge $+2/3$ quarks are $(6N+3)\times(6N+3)$ non-Hermitian matrices for three generations of quarks with associated KK excitations:

$$\left(\bar{q}_1^{(0)}, \bar{q}_2^{(0)}, \bar{q}_3^{(0)}, \bar{q}_1^{(1)}, \bar{q}_2^{(1)}, \bar{q}_3^{(1)}, \bar{u}_1^{(1)}, \bar{u}_2^{(1)}, \bar{u}_3^{(1)}, \ldots\right)_R M_u \left(u_1^{(0)}, u_2^{(0)}, u_3^{(0)}, q_1^{(1)}, q_2^{(1)}, q_3^{(1)}, u_1^{(1)}, u_2^{(1)}, u_3^{(1)}, \ldots\right)_L^T \quad (18)$$

where $\{\bar{q}_{iL}^{(0)}\}$, $\{\bar{q}_{iL}^{(n)}\}$, and $\{q_{iR}^{(n)}\}$ represent the charge $2/3$ components of the doublets. There are similar mass matrices $M_d$ and $M_d^\dagger$ for the charge $-1/3$ quarks:

$$\left(\bar{q}_1^{(0)}, \bar{q}_2^{(0)}, \bar{q}_3^{(0)}, \bar{q}_1^{(1)}, \bar{q}_2^{(1)}, \bar{q}_3^{(1)}, \bar{d}_1^{(1)}, \bar{d}_2^{(1)}, \bar{d}_3^{(1)}, \ldots\right)_R M_d \left(d_1^{(0)}, d_2^{(0)}, d_3^{(0)}, q_1^{(1)}, q_2^{(1)}, q_3^{(1)}, d_1^{(1)}, d_2^{(1)}, d_3^{(1)}, \ldots\right)_L^T \quad (19)$$

where $\{\bar{q}_{iL}^{(0)}\}$, $\{\bar{q}_{iL}^{(n)}\}$, and $\{q_{iR}^{(n)}\}$ now represent the charge $-1/3$ components of the doublets. Observe that we denote the charge $2/3$ quark mixing matrix as $M_u$ when we are considering inter-generational mixing, and as $M_t$ when we are reducing our attention to the $3\times 3$ top quark mixing.

The charge $2/3$ mass eigenstates $\{u, c, t, u_1^\bullet, c_1^\bullet, t_1^\bullet, u_1^\circ, c_1^\circ, t_1^\circ, \ldots\}$ are linear superpositions of the weak eigenstates $\{u_1^{(0)}, u_2^{(0)}, u_3^{(0)}, q_1^{(1)}, q_2^{(1)}, q_3^{(1)}, u_1^{(1)}, u_2^{(1)}, u_3^{(1)}, \ldots\}$:

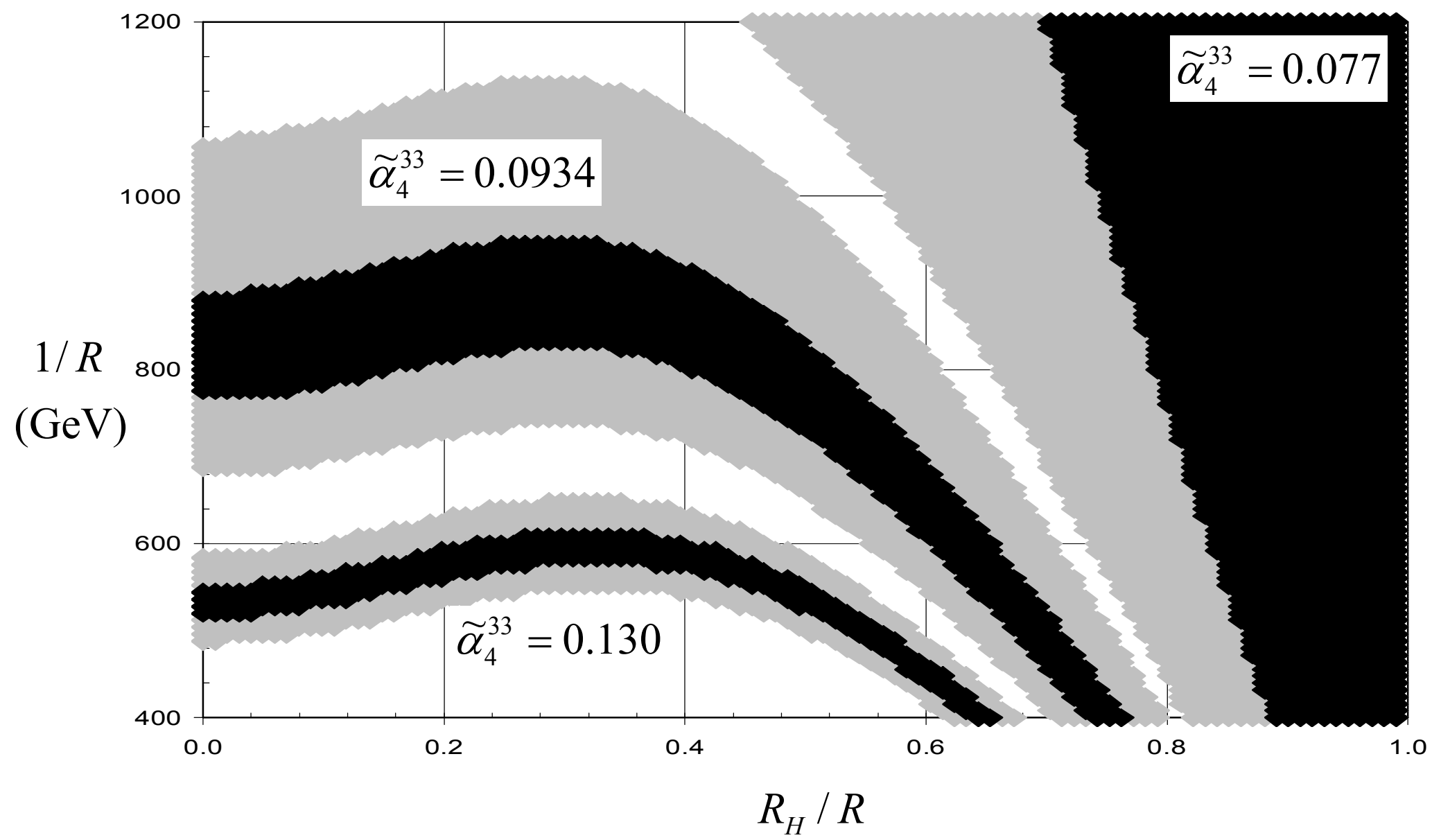

**Fig. 3**. The allowed ranges of the ratio $R_H/R$ and quark compactification scale $1/R$ are depicted for the case $0 < R_H < R$ at the 1-sigma (♦) and 3-sigma (♦) level for three different values of $\tilde{\alpha}_4^{33}$, using $m_t = 171.4 \pm 2.1\,\mathrm{GeV}$.

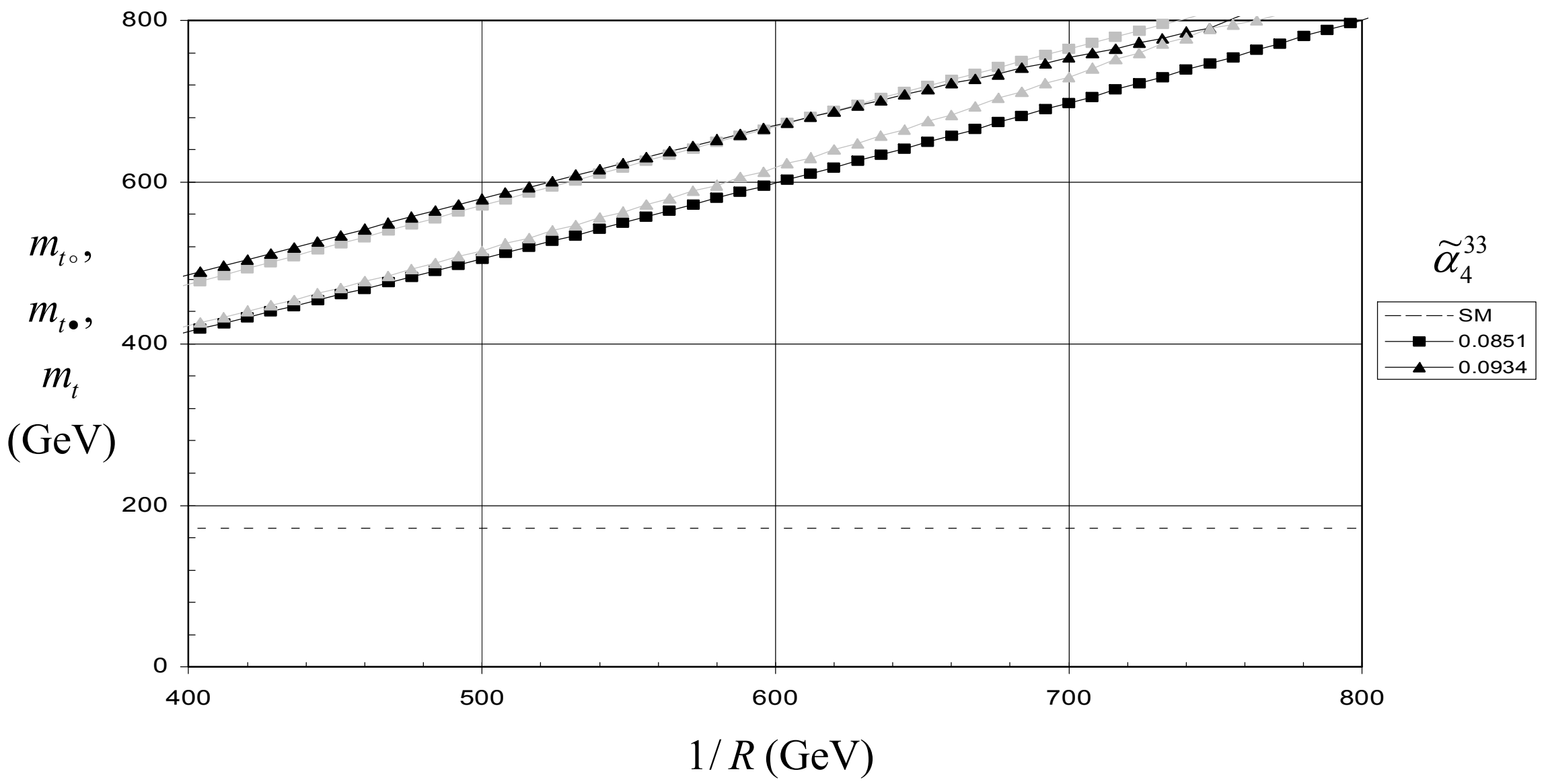

**Fig. 4.** The mass eigenvalues $m_t, m_{t_1^\bullet}$, and $m_{t_1^\circ}$ of the top quark mass matrix $M_t$ are illustrated as functions of the quark compactification scale $1/R$ for the central allowed values of $R_H/R$ from Fig. 3 for two different values of $\tilde{\alpha}_4^{33}$. The black (▲) and gray (▲) curves illustrate the mass splitting between $m_{t_1^\bullet}$ and $m_{t_1^\circ}$. The dashed line (– – –) depicts the observed top quark mass $m_t = 171.4 \pm 2.1\,\mathrm{GeV}$.

$$\left(u,c,t,u_1^\bullet,c_1^\bullet,t_1^\bullet,u_1^\circ,c_1^\circ,t_1^\circ,\ldots\right)_L^T = U_L^{-1}\left(q_1^{(0)},q_2^{(0)},q_3^{(0)},q_1^{(1)},q_2^{(1)},q_3^{(1)},u_1^{(1)},u_2^{(1)},u_3^{(1)},\ldots\right)_L^T$$
$$\left(u,c,t,u_1^\bullet,c_1^\bullet,t_1^\bullet,u_1^\circ,c_1^\circ,t_1^\circ,\ldots\right)_R^T = U_R^{-1}\left(u_1^{(0)},u_2^{(0)},u_3^{(0)},q_1^{(1)},q_2^{(1)},q_3^{(1)},u_1^{(1)},u_2^{(1)},u_3^{(1)},\ldots\right)_R^T \qquad (20)$$

and similarly for the charge $-1/3$ mass eigenstates:

$$\left(d,s,b,d_1^\bullet,s_1^\bullet,b_1^\bullet,d_1^\circ,s_1^\circ,b_1^\circ,\ldots\right)_L^T = D_L^{-1}\left(q_1^{(0)},q_2^{(0)},q_3^{(0)},q_1^{(1)},q_2^{(1)},q_3^{(1)},d_1^{(1)},d_2^{(1)},d_3^{(1)},\ldots\right)_L^T$$
$$\left(d,s,b,d_1^\bullet,s_1^\bullet,b_1^\bullet,d_1^\circ,s_1^\circ,b_1^\circ,\ldots\right)_R^T = D_R^{-1}\left(d_1^{(0)},d_2^{(0)},d_3^{(0)},q_1^{(1)},q_2^{(1)},q_3^{(1)},d_1^{(1)},d_2^{(1)},d_3^{(1)},\ldots\right)_R^T \qquad (21)$$

The quark mass matrices are diagonalized via bi-unitary transformations:

$$U_L^{-1} M_u U_R = M_u^{diag} \quad , \quad D_L^{-1} M_d D_R = M_d^{diag} \qquad (22)$$

The mass eigenstates mix via the $(6N+3)\times(6N+3)$ generalization of the usual $3\times 3$ Kobayashi-Maskawa (KM) matrix [9]:

$$V = U_L^\dagger D_L \qquad (23)$$

## 4. Collider Phenomenology

One consequence of the inter-mode quark mixing in the non-universal Higgs model is that the KK excitations of the quarks can decay to a Higgs and a zero-mode quark at the tree-level through the Yuwawa interactions. This is in striking contrast to the UED case, in which the KK excitations are stable at the tree-level as a result of conservation of KK number. In particular, the physical $n=1$ level KK excitations of the top quark, $t_1^\bullet$ and $t_1^\circ$, can decay to the observed $n=0$ top quark $t$ and a Higgs:

$$t_1^\bullet \to t + H \quad , \quad t_1^\circ \to t + H \qquad (24)$$

Such couplings arise because the mass matrix and the Yukawa coupling matrix are not proportional. As noted earlier, while the mass matrix is given by Eq. 9, the corresponding Yukawa coupling matrix for the fields does not have the $\pm 1/R$ terms in the diagonal $(2,2)$ and $(3,3)$ elements, which arise from the 5D kinetic terms. The effective 4D Yukawa coupling matrix can be written as

$$U_L^{-1} Y U_R = U_L^{-1} M_u U_R - U_L^{-1} K U_R = M_u^{diag} - U_L^{-1} K U_R \qquad (25)$$

where

$$K \equiv \begin{pmatrix} 0 & 0 & 0 \\ 0 & \pm 1/R & 0 \\ 0 & 0 & \mp 1/R \end{pmatrix} \qquad (26)$$

Thus, the vertex factors for the $t_1^{\bullet} tH$ and $t_1^{\circ} tH$ couplings are obtained from the appropriate matrix elements of $U_L^{-1} K U_R$, which we denote by $Y_{21}$ and $Y_{31}$, respectively.

One effect of the inter-mode quark mixing is a significant splitting in the masses of $t_1^{\bullet}$ and $t_1^{\circ}$. In the UED model, these masses are degenerate and equal to $\sqrt{m_t^2 + 1/R^2}$. In the non-universal Higgs model, one of the $n = 1$ level KK excitations becomes significantly lighter than $\sqrt{m_t^2 + 1/R^2}$, while the other becomes significantly heavier. For example, in the case $R_H = 0$, tree-level mixing between $u_3^{(0)}$, $u_3^{(1)}$, and $q_3^{(1)}$ results in a lighter $n = 1$ mass eigenvalue of $m_{t_1}^{light} \approx 1/R - \upsilon/2$. Which is heavier – $t_1^{\bullet}$ or $t_1^{\circ}$ – depends on which $1/R$ term is positive and which is negative in Eq. 26. We henceforth refer to the two $n = 1$ top quarks as $t_1^{light}$ and $t_1^{heavy}$ when a distinction is needed.

The lighter $n = 1$ top quark decays predominantly via $t_1^{light} \to t + H$, whereas the $t_1^{heavy}$ can have additional decay modes depending on the KK spectrum. If $1/R$ exceeds $\approx m_H + m_t + \upsilon/2 \approx m_H + 295\,\text{GeV}$,‡ then a $t_1^{light}$ produced at a high-energy collider will decay into a $t$ and an $H$ within the detector. Otherwise, the decay $t_1^{light} \to t + H$ is kinematically suppressed; however, in this case the production of $t_1^{heavy}$ will still contribute toward Higgs production.

Thus, the non-universal model may serve as a significant source of Higgs production at the LHC. The direct production of $t_1^{light} \bar{t}_1^{light}$ will result in $t\bar{t}HH$ production through subsequent decays of $t_1^{light}$ and $\bar{t}_1^{light}$. The pair of Higgs bosons along with $t\bar{t}$ will help distinguish the non-universal Higgs model from the SM background. The primary SM contributions to Higgs production at the LHC include gluon fusion $(gg \to H)$, weak boson fusion $(qq \to qqH)$, $W^{\pm}$ and $Z$ boson associated production ($q\bar{q} \to ZH$ and $q\bar{q} \to Wh$), bottom fusion $(b\bar{b} \to H)$, and associated top quark production ($gg \to t\bar{t}H$ and $q\bar{q} \to t\bar{t}H$).

We compute the cross section for $t_1^{light} \bar{t}_1^{light}$ production at the LHC and the results are displayed in Fig. 5 as a function of the compactification scale $1/R$ for $R_H / R = 0$, ¼, and ½. The uncertainties in the tree-level cross section may be as high as approximately fifty percent owing to uncertainties in the CTEQ [10] parton distributions, the scale dependence $Q$ in the strong coupling and parton distributions, and the neglect of higher-order corrections. In light of these uncertainties, we conservatively require a minimum of

‡ This conservative limit is based on the case of maximal mixing – i.e. $R_H = 0$. The limit is relaxed somewhat as $R_H / R$ increases.

100 events observed (with the anticipated luminosity of 100 $\text{fb}^{-1}$) and a KK signal at least comparable to SM Higgs production in placing our bounds.

As can be seen from Fig. 5, the KK contribution to Higgs production in the non-universal Higgs model turns out to be quite significant, and can be as large as 60 pb for a compactification scale of 400 GeV. Thus, depending on the Higgs mass and the compactification scale, this mechanism can be the dominant source for Higgs boson production at the LHC. We also see that the KK contribution to the LHC cross section for Higgs production is comparable to the SM contributions to Higgs production of about 1-10 pb [11], depending on the Higgs mass, for compactification scales up to about 600 GeV. The SM cross section for double Higgs production is much smaller, about 10-50 fb [12]. The KK signal is comparable to the double Higgs background for compactification scales up to about 1.0-1.5 TeV. The final state signal – the top quark, anti-top quark, and the two Higgs bosons – is also quite unique, arising from the decays of the $t_1^{light}$ and $\bar{t}_1^{light}$.

The KK cross section depends largely upon the compactification scale $1/R$ and somewhat on the ratio $R_H/R$; however, it is independent of the Higgs mass, except for whether or not the Higgs is too heavy for the $t_1^{light} \rightarrow t + H$ decay to occur. Although the ratio $R_H/R$ does cause a discernible shift in the cross section in Fig. 5, its main role is whether or not there is just enough mixing for the $t_1^{light} \to t + H$ decay channel to open (which is significant if $1/R$ is near its lower limit of about 350-400 GeV or if the Higgs mass is rather large). There is a lower bound on $1/R$ in Fig. 5 for $R_H/R$ equal to ¼ and ½. The reason for this lower bound can be understood by comparing with Fig. 3 – i.e. small values of $1/R$ are inaccessible (within the $3\sigma$ tolerance) for any value of $\tilde{\alpha}_4^{33}$ for values of $R_H/R$ from about 0.1 to 0.5.

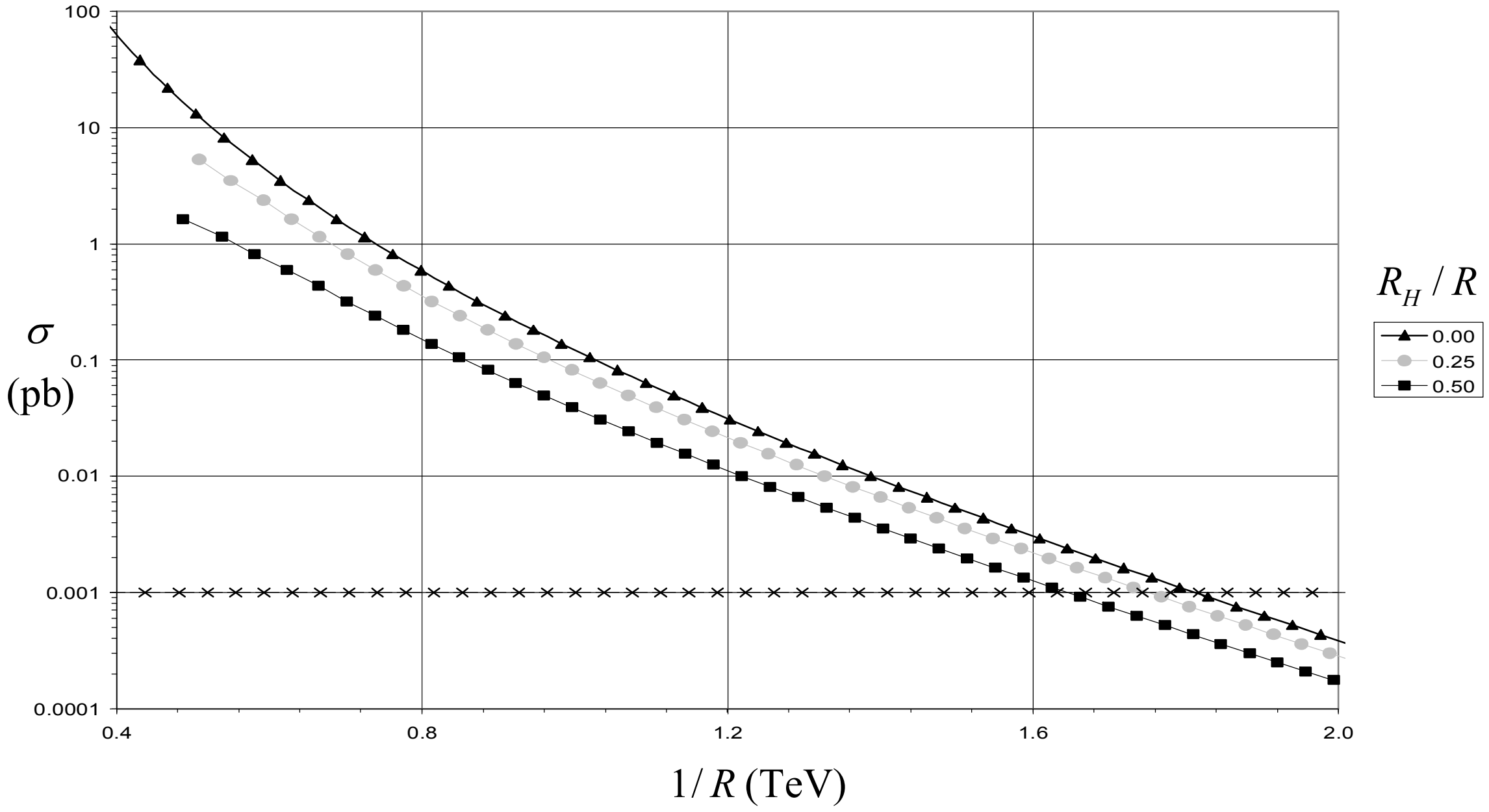

**Fig. 5.** The cross section for $t_1^{\bullet}\bar{t}_1^{\bullet}$ or $t_1^{\circ}\bar{t}_1^{\circ}$ production (whichever is lighter) at the LHC is illustrated as a function of the quark compactification scale $1/R$ for the central allowed values of $\tilde{\alpha}_4^{33}$ from Figs. 1 and 3 for three different values of $R_H/R$. The ×'s at 0.001 pb represent 100 events at the LHC based on the projected luminosity of 100 $\text{fb}^{-1}$.

## 5. Conclusions

In this work, we have examined inter-mode quark mixing in the non-universal Higgs model, where the Higgs is wholly or partially excluded from propagating into one $\text{TeV}^{-1}$-size OUED. We find that tree-level KK number conservation is violated in the non-universal Higgs model, which results in inter-mode mixing among the quarks through KK number violation in the Yukawa interactions. In particular, the zero modes of the quarks mix with their lowest-lying associated KK excitations. This is in contrast to the UED scenario, in which KK number conservation prohibits inter-mode mixing. We analyze in detail the effects resulting from tree-level mixing among the $n = 0$ and $n = 1$ modes in the top quark sector where this mixing is large.

In the non-universal Higgs model, the physical quarks $\{t, t_1^\bullet, t_1^\circ\}$ are a linear combination of the weak eigenstates $\{q_3^{(0)}, q_3^{(1)}, u_3^{(1)}\}$. This $3 \times 3$ mixing involves just three parameters: $\tilde{y}_4^{33}$, $R_H/R$, and $1/R$. There is one additional parameter involved in the bottom sector: $y_4^{33}$. The lighter generations and $n \geq 2$ modes approximately decouple – i.e. inter-mode quark mixing is most significant for the top quark. Since the observed top quark is actually a combination of $u_3^{(0)}$, $u_3^{(1)}$, and $q_3^{(1)}$, the lightest third-generation charge $2/3$ quark mass eigenvalue is constrained to correspond to the measured top quark mass (171.4 ± 2.1 GeV). We computed the effects that the model parameters have on this observable and established the allowed region of parameter space.

As a result of the inter-mode quark mixing, the KK excitations of the quarks can decay to a Higgs boson and a corresponding zero-mode quark. In particular, the lighter of the two $n = 1$ KK excitations of the top quark – either the $t_1^\bullet$ or $t_1^\circ$, depending on the signs of the contributions of the kinetic terms to the top quark mass matrix – decays predominantly to $tH$. We computed the LHC cross section for $t_1^{light}\bar{t}_1^{light}$ production, where $t_1^{light}$ is the lighter of $t_1^\bullet$ and $t_1^\circ$. The subsequent decays of $t_1^\bullet$ and $t_1^\circ$ to $tH$ (which is the dominant decay mode) give rise to $t\bar{t}HH$ final states, and the ensuing signals can be detected at the LHC. The effect that the non-universal Higgs has on Higgs production at the LHC turns out to be quite large: The KK cross section is comparable to the SM contribution to Higgs production for compactification scales up to about 600 GeV, and is even larger for smaller values of the compactification scale. Thus, this mechanism can be the dominant source for the Higgs production at the LHC. Furthermore, the two Higgs plus $t\bar{t}$ signal is distinct from the primary sources of Higgs production in the SM, which is either a single Higgs boson, or a single Higgs boson in association with other SM particles.

The research of SN is supported by the US Department of Energy Grant numbers DE-FG02-04ER41306 and DE-FG02-04ER46140. CM would like to thank OSU for its kind hospitality during this research.